# Light Dragging, the Origin of Hubble's Constant

## Abstract

**An explanation for nebulae redshift is presented based on spacetime dragging of light**

## I. Introduction

Recently E. Harrison has argued[1] the redshift distance law proposed by Hubble[2,3] and the velocity distance law developed later on theoretical grounds has no general proof demonstrating the two laws are actually equivalent. It is the purpose of this paper to account for the nebular redshift law of Hubble based up two principles:

**1)** Spacetime motion and light dragging[4].
**2)** An overall spacetime index of refraction based on Hubble's constant.

## II. Development

In 1929 Hubble presented the case for a linear relationship between redshift z and extragalactic nebulae distance[5]:

$$z = \text{constant} \times L \qquad (2.1)$$

This linear redshift-distance z(L) is usually expressed in the form as:

$$zc = H_0 L \qquad (2.2)$$

Here c is the speed of light; $H_0$ is Hubble's constant, which ranges somewhere near **100 km/s Mpc$^{-1}$**; L is the distance to some galaxy to which the redshift z is being measured; the redshift is defined by:

$$z = \frac{\lambda - \lambda_0}{\lambda_0} \qquad (2.3)$$

An alternative form of Hubble's relationship is given by:

$$v_r = H_0 L = zc \qquad (2.4)$$

where $v_r$ the radial velocity of the distant Nebulae [Note $v_r$ is positive for recessional velocity].

## III. Motion of Spacetime and Hubble's Constant

Motion of spacetime and its effects on matter is not a new concept[6, 7, 8, 9, 10]. For example NASA has launched the gravity probe B[11] to measure spacetime drag (Thirring Effect)[12] caused by the rotation of the Earth. In 2007 Christensen[13] proposed a spacetime model based on normal coordinates and oscillating gravitons. It was shown that islands of spacetime rotation could form in the gravitational field having angular frequency of:

$$\Omega = 1.7 \times 10^{-11} \text{ rad/sec} \tag{3.1}$$

This value is in close agreement with Gödel's rotating universe model having angular velocity[14] of $10^{-11}$ rad/sec, and Bayin's of $10^{-13}$ rad/sec. Bayin's metric is given by:

$$ds^2 = dt^2 - e^{g(t)}\left(\frac{dr^2}{1-r^2/R^2} + r^2 d\Omega^2\right) + 2r^2 \sin^2\theta e^g \Omega(r,t) d\phi dt \tag{3.2}$$

where $\Omega(r, t)$ is the metric rotation function related to the local dragging of inertial frames.

If spacetime indeed experiences various motions[15, 16, 17] then it would be capable of dragging light along with it[18, 19]. Under such conditions spacetime could be treated as a kind of "gravitomagnetic medium", hence the work of Fermi[20], Jones[21] and Player[22] could be applied to explain the nebular redshift observed by Hubble and others.

## IV. Maxwell's Equations for a Rotating Medium Dragging Light

Player showed for an electric field in a rotating medium:

$$\nabla^2 \vec{D} = \frac{\varepsilon}{c^2} \frac{\partial^2 \vec{D}}{\partial t^2} - 2\vec{\omega} \times \frac{\partial \vec{D}}{\partial t} \tag{4.1}$$

To order v/c **D** has the form:

$$\vec{D} = \vec{D}_0 e^{j(\omega t - kx)} \tag{4.2}$$

where

$$\vec{D}_0 = D_{0x}\hat{x} + D_{0y}\hat{y} \tag{4.3}$$

The relationship between the components of the electric field may be investigated by substituting **D** into the differential equation (3.2) whereupon it yields:

$$k^2 D_{0x} = \frac{\varepsilon}{c^2}\omega^2 D_{0x} - 2j\omega\Omega\frac{\varepsilon-1}{c^2} D_{0y} \tag{4.4}$$

$$k^2 D_{0y} = \frac{\varepsilon}{c^2}\omega^2 D_{0y} - 2j\omega\Omega\frac{\varepsilon-1}{c^2}D_{0x} \qquad (4.5)$$

Which admit of the following two solutions: a right-handed circularly polarized wave (in which the vectors at any point rotate clockwise when viewed against the direction of propagation):

$$D_{0y} = jD_{0x} \qquad (4.6)$$

$$k^2 = \frac{\varepsilon}{c^2}\omega^2 - 2j\omega\Omega\frac{\varepsilon-1}{c^2} \qquad (4.7)$$

For the right-handed wave, in case of positive Ω, the effective frequency ω' is increased since the electric vector rotates in the sense opposite to that of the rotation of the "medium" (in the case here spacetime); thus from the above expressions, and using the relation n = ck/ω, there are refractive indices associated with the waves of each handedness:

$$n_R = n(\omega+\Omega)\left[n(\omega+\Omega) - \frac{1}{n(\omega+\Omega)}\right]\frac{\Omega}{\omega} \qquad (4.8)$$

$$n_R = n(\omega-\Omega)\left[n(\omega-\Omega) - \frac{1}{n(\omega-\Omega)}\right]\frac{\Omega}{\omega} \qquad (4.9)$$

The "Effective Power of Rotation" of the medium (or phase difference per unit length), and is given by:

$$\delta(\omega) = (n_R - n_L)\frac{\pi v}{c} \qquad (4.10)$$

Retaining only terms linear in Ω one easily obtains:

$$\delta(\omega) = \left(v\frac{dn}{dv} + n - \frac{1}{n}\right)\frac{\Omega}{c} \qquad (4.11)$$

where n now denotes the refractive index at the rest frame frequency, or

$$\delta = \left(n_g - \frac{1}{n_\phi}\right)\frac{\Omega}{c} \qquad (4.12)$$

Using the "group refractive index" $n_g = n + \upsilon dn/d\upsilon$ and distinguishing the phase index by its own subscript φ. However this result has been derived for a single medium undergoing rotation, whereas the spacetime model presented here is comprised of islands of space time rotation; each on contributing to light dragging, hence to the redshift of nebulae light. However, to account for all spacetime motion inducing light dragging, an index of refraction $N_s$ for spacetime is defined to be:

$$N_s \equiv \left( n_g - \frac{1}{n_\phi} + n_m \right) \tag{4.13}$$

where $n_m$ represents the index of refraction effects from all other spacetime motions.

Letting L represent the distance to an observed galaxy or celestial object, the associated redshift z is then directly related to the spacetime index of refraction given by:

$$\delta(\omega) Lc = zc = N_s \frac{\Omega}{2} L = HL \tag{4.14}$$

Thus Hubble's constant is based on the index of refraction of spacetime[23] and the graviton frequency:

$$H = N_s \frac{\Omega}{2} \tag{4.15}$$

From this relationship the index of refraction for spacetime is given to be:

$$N_s = \frac{2H}{\Omega} = \frac{2(100 km/s)}{(1.7x10^{-11}/s)Mpc} \frac{Mpc}{(3.085x10^{22} meters)} = 3.8x10^{-7} \tag{4.16}$$

Note that the index of refraction must be related to the spacetime's relative permittivity $\varepsilon_r$ and relative permittivity $\mu_r$:

$$N_s = \sqrt{\mu_r \varepsilon_r}$$

**IV. Conclusion**

It was shown that if spacetime undergoes rotation, an overall index of refraction can be associated with spacetime that is directly related to Hubble's constant. This refraction constant provides for an explanation for nebulae redshift.

---

[1] E. Harrison. "The Red-Shift Distance and Velocity-Distance Laws." ApJ. **403**:p28 (1993)